\documentstyle[epsf]{article}                                              %
\font\tenrm=cmr10
\font\tenit=cmti10
\font\elevenbf=cmbx10 scaled\magstep 1
\font\elevenrm=cmr10 scaled\magstep 1
\font\elevenit=cmti10 scaled\magstep 1

\font\ninerm=cmr9

\def\vu{\varepsilon}
\def\a{\alpha}
\def\b{\beta}
\def\c{\chi}
\def\d{\delta}
\def\f{\phi}                    %       \varphi

\def\h{\eta}

\def\k{\kappa}                  % Also, \varkappa (see below)
\def\l{\lambda}
\def\m{\mu}
\def\n{\nu}

\def\p{\pi}                     % Also, \varpi
\def\t{\tau}

\def\D{\Delta}
\def\F{\Phi}
\def\G{\Gamma}

\def\L{\Lambda}

\def\P{\Pi}

\def\S{\Sigma}

\def\cs{{\cal S}}

\newcommand{\extraspace}{\addtolength{\abovedisplayskip}{2mm}
                        \addtolength{\belowdisplayskip}{2mm}
                        \addtolength{\abovedisplayshortskip}{2mm}
                        \addtolength{\belowdisplayshortskip}{2mm}}
\newcommand{\be}{\begin{equation}\extraspace}

\newcommand{\ee}{\end{equation}}

\newcommand{\bea}{\begin{eqnarray}\extraspace}
\newcommand{\beastar}{\begin{eqnarray*}\extraspace}
\newcommand{\eea}{\end{eqnarray}}
\newcommand{\eeastar}{\end{eqnarray*}}

\newcommand{\nonu}{\nonumber \\[2mm]}

\newcommand{\dis}{\displaystyle}

\newcommand{\half}{\frac{1}{2}}

\newcommand{\var}{\varphi}

\newcommand{\del}{\partial}

\newcommand{\slt}{sl(2)}

\newcommand{\np}{Nucl.\ Phys.\ }

\newcommand{\pl}{Phys.\ Lett.\ }

\renewenvironment{thebibliography}[1]
 { \elevenrm
   \begin{list}{\arabic{enumi}.}
    {\usecounter{enumi} \setlength{\parsep}{0pt}
     \setlength{\itemsep}{3pt} \settowidth{\labelwidth}{#1.}
     \sloppy
    }}{\end{list}}

\begin{document}
\vglue -0.4cm
\noindent June 1993 \hfill LBL-34240, UCB-PTH-93/21\\
\begin{center}
\vglue -0.4cm
{{\elevenbf        \vglue 10pt
               EXTENSIONS OF 2D GRAVITY\footnote{\ninerm\baselineskip=11pt This
work was
supported in part by the Director,
Office of Energy Research, Office of High Energy and Nuclear Physics,
Division of High Energy Physics of the U.S. Department of Energy
under Contract DE-AC03-76SF00098 and in part by the National Science
Foundation under grant PHY90-21139.
}
\footnote{\ninerm Invited talk at the ``{\it Journ\'ees Relativistes}'',
Brussels, April, 1993.}
\\}
\vglue 0.8cm
{\tenrm Alexander Sevrin\footnote{\ninerm\baselineskip=11pt Address after
October 1st, 1993: CERN, TH Division, CH-1211 Geneva, Switzerland.}\\}
\baselineskip=13pt
{\tenit Department of Physics, University of California at Berkeley \\}
{\tenit and \\}
{\tenit Theoretical Physics Group, Lawrence Berkeley Laboratory, \\}
\baselineskip=12pt
{\tenit Berkeley, CA 94720, USA\\}}

%\vglue 0.3cm
%{\tenrm and\\}
%\vglue 0.3cm
%{\tenrm SECOND AUTHOR'S NAME\\}
%{\tenit Group, Company, Address, City, State ZIP/Zone, Country\\}
%\vglue 0.8cm
%{\tenrm ABSTRACT}

\end{center}

\vglue 0.3cm
{\rightskip=3pc
\leftskip=3pc
\tenrm\baselineskip=12pt
\noindent
After reviewing some aspects of gravity in two dimensions, I show that
non-trivial embeddings of $sl(2)$ in a semi-simple (super) Lie algebra give
rise to a very large class of extensions of 2D gravity. The induced action is
constructed as a gauged WZW model and an exact expression for the effective
action is given.}
\setcounter{section}{1}
\setcounter{equation}{0}
\vglue 0.6cm
{\elevenbf\noindent 1. Introduction}
\vglue 0.4cm
\baselineskip=14pt
\elevenrm
Conformally invariant theories in two dimensions play an important role in the
study of string theories, second order phase transitions and integrable
systems. The Virasoro algebra and its extensions${}^1$ form the cornerstone of
conformal field theories. More recently a lot of attention was devoted to the
study of a particular class of conformal field theories: gravity in two
dimensions. As I will show in the next section, gravity in two dimensions is a
purely quantum mechanical artefact. Contrary to the case of higher dimensions,
gravity in two dimensions allows for infinitely many extensions such as higher
spin fields, supersymmetry, Yang-Mills symmetries, etc. These extensions are in
one to one correspondence with the extensions of the Virasoro algebra.

The interest in 2D gravity arose from its close relation to non-critical string
theories. In these theories, the matter sector is a minimal model. A
propagating gravity sector contributes to the conformal anomaly in such a way
that the conformal anomaly of the combined matter-gravity-ghost sectors
vanishes. Non-critical string theories allow for a non-perturbative treatment
thus providing a testing ground for techniques and ideas which might be
applicable to more ``realistic'' string theories${}^2$. More recently 2D
gravity gave rise to models which made the study of quantum aspects of black
hole evaporation possible${}^3$.
%\newpage
\setcounter{section}{2}
\setcounter{equation}{0}
\vglue 0.4cm
{\elevenbf\noindent 2. Gravity in Two Dimensions}
\vglue 0.4cm
\baselineskip=14pt
\elevenrm
The Einstein-Hilbert action in two dimensions
\bea
S_{\rm EH}=\frac{1}{4\pi}\int_\S d^2x\sqrt{g}R^{(2)},
\eea
has no dynamical content as it gives the Euler characteristic of the surface
$\S$: $\c (\S)=2 -2 h$, a topological invariant. Consider a scalar field $\f$
coupled to gravity:
\bea
S_{\rm SF}=\frac{1}{4\pi}\int d^2x\sqrt{g}g^{\mu\nu}\partial_\mu\phi
\partial_\nu\phi.\label{s1}
\eea
Introducing light-cone coordinates, one parametrizes the metric as
$g_{+-}=e^\var$, $g_{\pm\pm}=2e^\var\m_{\pm\pm}$ and Eq. (\ref{s1}) gets
rewritten as
\bea
S_{\rm SF}=\frac{1}{2\pi}\int d^2x\frac{1}{\sqrt{1-4\m_{++}\m_{--}}}
\left(\del_+\f\del_- \f+2\m_{++}T_{--}(\f) + 2\m_{--}T_{++}(\f)\right),
\label{s2}
\eea
where the energy-momentum tensor is given by $T_{\pm\pm}(\f)=-\frac 1 2
\del_\pm \f \del_\pm\f$. The modes of the energy-momentum tensor form two
commuting copies of the Virasoro algebra with central extension $c=1$. As the
action has no explicit dependence on the conformal mode $\var$ anymore, the
theory is not only invariant under general coordinate transformations but under
local Weyl rescalings of the metric as well. There are as many gauge symmetries
as gauge fields so one concludes that here also there is no (except for moduli)
gravitational content.

Quantum mechanically, one of the symmetries becomes anomalous and the metric
acquires dynamics. To see this, one passes to the light-cone gauge
$\var=\m_{++}=0$ and the action Eq. (\ref{s2}) reads now:
\bea
S_{\rm SF}' = \frac{1}{2\p}\int( \del_+\f\del_-\f + 2\m_{--}T_{++}(\f)).
\eea
Classically there is still one gauge symmetry left:
\bea
\d\f&=&\vu_-\del_+\f,\nonu
\d\m_{--}&=&\del_-\vu_-+\vu_-\del_+\m_{--}-\del_+\vu_-\m_{--},\label{gs1}
\eea
with $\vu_-$ an arbitrary infinitesimal parameter. The induced action $\G
[\m_{--} ]$, is defined by
\bea
e^{\dis - \G [\m_{--} ]}=\int [d\f ] e^{\dis - S_{\rm SF}'(\f, \m_{--})
}=\left< e^{\dis -\frac 1 \p \int \m_{--}T_{++}}\right>.\label{induced}
\eea
In diagrams $\G [\m_{--} ]$ is given by:

\epsffile{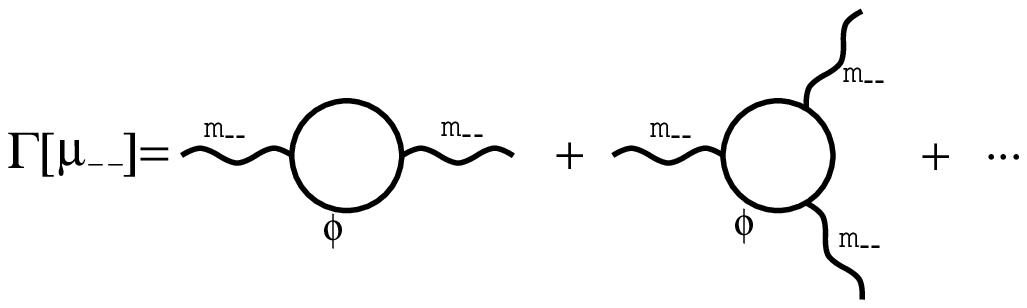}
\smallskip

\noindent If the symmetry Eq. (\ref{gs1}) were to  persist at quantum level,
the induced action would vanish. However one easily shows that

\epsffile{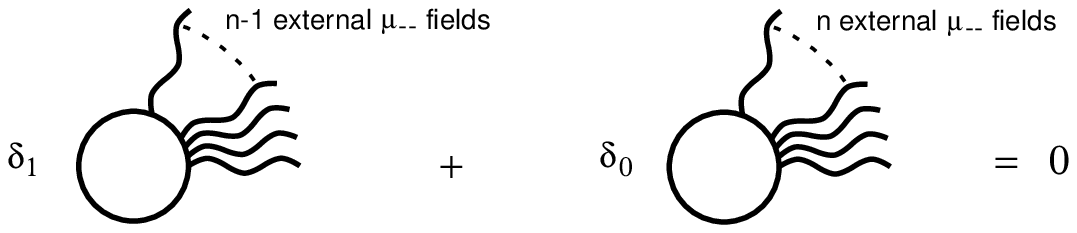}
\smallskip

\noindent where $\d_0\m_{--}=\del_-\vu_-$ and $\d_1\m_{--}=
\vu_-\del_+\m_{--}-\del_+\vu_- \m_{--}$. So the anomaly is given by the $\d_0$
variation of the two-point diagram:

\medskip
\epsffile{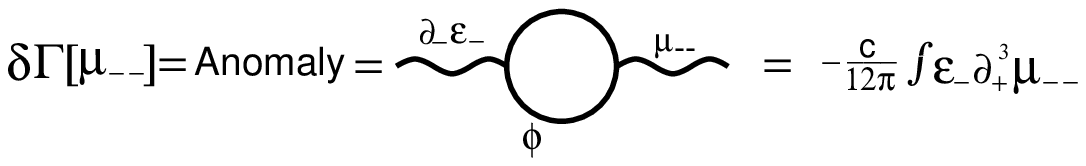}
\smallskip

\noindent where in our case $c=1$. Having computed the anomaly, the Ward
identity for $\G [\m_{--} ]$ follows:
\bea
\left( \del_--2\del_+\m_{--}-\m_{--}\del_+\right)\frac{\d \G [\m_{--}
]}{\d\m_{--}}=\frac{c}{12\p}\del_+^3\m_{--}.
\eea
This gives a functional differential equation for $\G [\m_{--} ]$. Methods to
solve this equation have been developed${}^{4,5}$:
\bea
\G [\m_{--} ]=\frac{c}{24\p}\int\del^2_+\m_{--} \frac{1}{\del_-}
\frac{1}{1-\m_{--} \frac{ \del_+}{\del_-}} \frac{1}{\del_+} \del_+^2\m_{--}.
\eea
Covariantizing this, one obtains the familiar result${}^{6}$
\bea
\G_{\rm cov}=\G [\m_{--} ]+\G [\m_{++} ]+\D [\m_{--},\m_{++},\var
]=\frac{c}{96\p}\int \sqrt{g} R \frac{1}{\Box} R,
\eea
which is manifestly invariant under general coordinate transformations. Adding
a cosmological constant to this and coupling it in a diffeomorphic invariant
way to a minimal model gives the action for a non-critical string. The
resulting model is  tuned such that the total central charge of the minimal
model and the gravity sector equals 26, precisely cancelling the contribution
to the central charge coming from the ghosts.

The effective action is obtained from the induced action by integrating over
the Beltrami differential $\m_{--}$:
\bea
e^{\dis -W[ \check{T}_{++}]}=\int [d \m_{--}]e^{\dis -\G [\m_{--}
]+\frac{1}{\p} \int \m_{--}\check{T}_{++}}.
\eea
Introduce the classical effective action $W_{\rm cl}[ \check{T}_{++}]$, simply
given by the Legendre transform of $\G [\m_{--} ]$:
\bea
W_{\rm cl}[ \check{T}_{++}]=\min_{\{ \m_{--}\} }\left( \G [\m_{--}
]-\frac{1}{\p} \int \m_{--}\check{T}_{++}\right).
\eea
The full effective action is equal to the classical one up to multiplicative
renormalizations${}^{7}$:
\bea
W[ \check{T}_{++}]=Z_cW_{\rm cl}[ Z_T \check{T}_{++}],
\eea
where
\bea
Z_c&=&\frac{c+\sqrt{(c-1)(c-25)}-37}{2c},\nonu
Z_T&=&\frac{c}{12+cZ_c}.
\eea
In section four, I will give a general proof for this, not only valid for pure
$2D$ gravity but for a very large class of extensions of it as well.
\setcounter{section}{3}
\setcounter{equation}{0}
\vglue 0.5cm
{\elevenbf\noindent 3. Extensions of 2D Gravity}
\vglue 0.4cm
\baselineskip=14pt
\elevenrm
Taking $W_3$ gravity as an example is most instructive and illustrates all
complications of the general case. Consider two free scalar fields, $\phi_1$
and $\phi_2$. For convenience, I use a matrix notation:
\bea
\phi = \left( \begin{array}{ccc} \phi_1 & 0 & 0 \\ 0 & \phi_2-\phi_1 & 0 \\ 0 &
0 & -\phi_2 \end{array} \right).
\eea
The action
\bea
{\cal S}=\frac{1}{2\p} \int \, tr \del_+ \phi \del_- \phi,
\eea
is invariant under
\bea
\d \phi =\vu_- \del_+ \phi +\l_{--} \del_{+} \phi\del_+ \phi-\frac 1 3 \l_{--}
\,tr \left\{\del_+ \phi\del_+ \phi\right\}\label{wsym}
\eea
provided $\del_-\vu_-=\del_-\l_{--}=0$. The Noether currents associated to
these symmetries are denoted by $T_{++}$ and $W_{+++}$: $T_{++}\propto
tr\{\del_+ \phi\del_+ \phi \}$ and $W_{+++}\propto tr\{\del_+ \phi\del_+
\phi\del_+ \phi \}$.
The modes of the currents $T$ and $W$ satisfy (at quantum level) the $W_3$
algebra with $c=2$:
\bea
\lbrack L_m, L_n \rbrack &=& \frac{c}{12} m(m^2-1) \d_{m+n,0} + (m-n) L_{m+n}
\nonumber\\
\lbrack L_m, W_n \rbrack &=& (2m-n) W_{m+n} \nonumber\\
\lbrack W_m, W_n \rbrack &=& \frac{c}{360} m(m^2-1)(m^2-4)\d_{m+n,0}
+ (m-n) \Bigg\{ \frac{1}{15} (m+n+3)(m+n+2)\nonu
&&- \frac{1}{6} (m+2)(n+2)
\Bigg\}
L_{m+n}  + \b (m-n) \L_{m+n},
\label{six}
\eea
where $m,n\in{\bf Z}$, $\b = 16/(22+5c)$
and
\bea
\L_m = \sum_{n\in {\bf Z}} : L_{m-n} L_n : - \frac{3}{10} (m+3)(m+2) L_m.
\eea
The normal ordering prescription is given by
$:L_mL_n:=L_mL_n$ if $\ m\leq -2$ and $:L_mL_n:=L_nL_m$ if $\ m >-2$.
The fact that the commutator of two generators is expressed not only as a
linear combination of the generators but contains composites of the generators
as well is a generic feature shared by most extensions of the Virasoro algebra.

The symmetry Eq. (\ref{wsym}) can be gauged by a simple minimal
coupling${}^{8}$:
\bea
S_{\rm SF}= \frac{1}{2\p} \int \, tr \del_+ \phi \del_- \phi + \frac{1}{\p}
\int  \left(
\m_{--}  T_{++}(\f) + \n_{---} W_{+++}(\f) \right).
\eea
The action is invariant under arbitrary transformations Eq. (\ref{wsym}),
provided the gauge fields transform as
\bea
\d\m_{--}&=& \del_-\vu_-+ \vu_-\del_+\m_{--} - \del_+ \vu_-\m_{--}-2\left(
\l_{--}\del_+\n_{---}-\n_{---}\del_+\l_{--} \right) T_{++},\nonu
\d\n_{---} &=& \del_- \l_{--}+2\l_{--} \del_+\m_{--}-\del_+ \l_{--}\m_{--}+
\vu_- \del_+ \n_{---}-2\del_+\vu_- \n_{---}.\label{wtrsf}
\eea
It is not very hard to find the $W_3$ analogue of the covariant action${}^{9}$
Eq. (\ref{s2}). The main observation is that Eq. (\ref{s2}) can be linearized
through the introduction of auxiliary fields (termed nested covariant
derivatives${}^9$) $F_+$ and $F_-$:
\bea
S_{\rm SF}=\frac{1}{\pi}\int d^2x \left(-\frac 1 2 \del_+\f\del_-
\f-F_+F_-+F_+\del_-\f+F_-\del_+\f+\hat{\m}_{++}T_{--}(F_+) +
\hat{\m}_{--}T_{++}(F_-)\right), \label{s3}
\eea
which upon elimination of $F_\pm$ through their equations of motion and
identifying $\m_{\pm\pm}=(1+\hat{\m}_{++}\hat{\m}_{--})^{-1}
\hat{\m}_{\pm\pm}$, reduces to Eq. (\ref{s2}). The action Eq. (\ref{s3}) is
gauge invariant with gauge transformations
\bea
\d \f&=&\vu_-F_++\vu_+F_-,\nonu
\d F_\pm&=&\vu_\mp\del_\pm F_\pm+\del_\pm \vu_\mp F_\pm,\nonu
\d\hat{\m}_{\mp\mp}&=&\del_\mp\vu_\mp+\vu_\mp\del_\pm \hat{\m}_{\mp\mp} -
\del_\pm \vu_\mp \hat{\m}_{\mp\mp}.
\eea
What happened is that because $F_\pm$ only transform under $\vu_\mp$
transformations and not under $\vu_\pm$ transformations, the problem reduced to
two copies of the chiral case and minimal coupling ensured gauge invariance. It
is clear now that exactly the same procedure can be applied for the case of
$W_3$. The action reads now
\bea
S_{\rm SF}&=& -\frac{1}{\p} \int \, \left( \frac 1 2 tr \del_+ \phi \del_- \phi
+
tr F_+F_-- tr F_+\del_-\f - tr F_-\del_+\f\right) \nonu
&&+\frac{1}{\p} \int  \left(
\m_{--}  T_{++}(F_+) + \n_{---} W_{+++}(F_+) + \m_{++}  T_{--}(F_-) + \n_{+++}
W_{---}(F_-) \right).
\eea
However, the auxiliary fields do now appear through cubic order in the action,
which prohibits a second order formulation.

The chiral $W_3$ symmetry Eqs. (\ref{wsym}, \ref{wtrsf}) is anomalous at the
quantum level${}^{4,5}$.  The induced action in the light-cone gauge is defined
similar to Eq. (\ref{induced}):
\bea
e^{\dis -\G [\m_{--},\n_{---} ]} = \left< e^{\dis - \frac{1}{\p} \int  \left(
\m_{--}  T_{++}(\f) + \n_{---} W_{+++}(\f) \right)} \right> \ .
\eea
A careful treatment of the non-linearities in the $W_3$ algebra${}^4$ reveals
that
the Ward identities contain non-local, subleading in $1/c$ terms. This in its
turn implies a $1/c$ expansion for the induced action:
\bea
\G [\m_{--},\n_{---}]=\sum_{n\geq 0}c^{1-n}\G^{(n)}
[\m_{--},\n_{---}].\label{onec}
\eea
Only $\G^{(0)}[\m_{--},\n_{---}]$ has been obtained in a closed form${}^5$.
The effective action is defined by:
\bea
e^{ \dis-W[\check{T}_{++},\check{W}_{+++}]} = \int [d\m_{--}] [d\n_{---}] \,
  e^{\dis -\G[\m_{--},\n_{---}]+\frac{1}{\p} \int\left(\m_{--}\check{T}_{++}
+\n_{---} \check{W}_{+++}\right) } .
\eea
The classical action is the Legendre transform of the leading term of the
induced action Eq. (\ref{onec}):
\bea
W^{(0)}[\check{T}_{++},\check{W}_{+++}]=\min_{\{ \m_{--},\n_{---}\}}\left(
c\G^{(0)}[\m_{--},\n_{---}]- \frac{1}{\pi}\int \left( \m_{--}\, \check{T}_{++}
+  \n_{---} \, \check{W}_{+++} \right)\right).
\eea
Just as for pure gravity, the full effective action is, except for a coupling
constant and a wavefunction renormalization, equal to the classical
action${}^{10,11}$:
\bea
W[\check{T}_{++},\check{W}_{+++}] = Z_c\, W^{(0)} \left[ Z_T \check{T}_{++},
Z_W \check{W}_{+++} \right] ,
\eea
\setcounter{equation}{0}
\setcounter{section}{4}
\vglue 0.2cm
{\elevenbf\noindent 4. Extended Gravity from Gauged WZW Models}
\vglue 0.4cm
\baselineskip=14pt
\elevenrm
In this section I will present a unifying treatment of extensions of the
Virasoro algebra and the corresponding effective gravity theories${}^{12-14}$.
The principle underlying this approach is quite simple. Consider a matter
system, where I denote the matter fields collectively by
$\f$, with as action $S[\f ]$ and with a set of $n$ symmetry currents,
denoted by $T_i[\f ]$, $i\in\{ 1, \cdots, n\}$, which forms an extended
Virasoro algebra. The induced action in the light-cone gauge is defined by
\bea
e^{\dis -\G[\m ]}=\int[d\f]\mbox{e}^{\dis -S[ \f]-\frac 1 \p \int \m^i
T_i[\f ]},
\eea
where $\m_i$ are sources. Alternatively $\m_i$ can be viewed as chiral gauge
fields or generalized Beltrami differentials. The effective
action\footnote{\ninerm\baselineskip=11pt Strictly speaking this is the
generating functional for connected Greens functions of $\m$, which only
upon a Legendre transform becomes the generating functional of 1PI Greens
functions which is usually called the effective action.} is defined by:
\bea
e^{\dis -W[\check{T}]}=\int[d \m]\mbox{e}^{\dis -\G[\m]+\frac 1 \p \int \m^i
\check{T}_i}=\int [d\f]\d(T[\f ] -\check{T})\mbox{e}^{\dis -S[ \f]}.
\eea
In order to evaluate this integral, one has to compute the Jacobian for going
from $T[\var ]$ to $\var$. Though this is usually impossible, we will be able
to do it by realizing the ``matter'' sector, {\it i.e.} $S[\f ]$, by a WZW
model for which a chiral, solvable group is gauged. The possible choices for
the gauge group are determined by the inequivalent, non-trivial embeddings of
$sl(2)$ in the Lie algebra.

Consider
a (super) Lie algebra $\bar{g}$. Call the affine extension of $\bar{g}$ with
level $\k$:  $\hat{g}$. The affine algebra is realized by a Wess-Zumino-Witten
theory with action $\k S^-[g]$.  Given a nontrivial embedding of $\slt$ in
$\bar{g}$, the adjoint
representation of $\bar{g}$ branches into irreducible representations of
$\slt$ which allows us to write the generators of $\bar{g}$ as
$t_{(jm,\a_j)}$ where
$j\in \half{\bf N}$ labels the irreducible representation
of $\slt$, $m$ runs from $-j$ to $j$ and $\a_j$ counts the multiplicity of the
irreducible representation $j$ in the branching. The $\slt$ generators $e_\pm$
and $e_0$ where $e_\pm\equiv t_{(1\pm 1,0)}/\sqrt{2}$ and $e_0\equiv
t_{(10,0)}$, satisfy the standard commutation relations: $[e_0,e_\pm ]=\pm 2
e_\pm$ and
$[e_+,e_-]=e_0$. The action of the $\slt$ algebra on the other generators is
given by
\bea
[e_0,t_{(jm,\a_j)}]&=&2m\, t_{(jm,\a_j)},\nonu
[e_\pm, t_{(jm,\a_j)}]&=&(-)^{j+m-\frac 1 2 \pm\frac 1 2}\sqrt{(j\mp m)(j\pm
m+1)} t_{(jm\pm 1,\a_j)}.
\eea
The $\slt$ embedding introduces a natural grading on $\bar{g}$ given by the
eigenvalue of $e_0$. I use the projection operators $\P$ to project Lie algebra
valued fields onto certain subsets of the $\slt$ grading, {\it e.g.}
$\P_+\bar{g}=\{ t_{(jm,\a_j)}|m > 0 ;\forall j,\a_j\}$,
$\P_{\geq m}\bar{g}=\{ t_{(jn,\a_j)}|n \geq m ;\forall j,\a_j\}$,
$\P_m\bar{g}=\{ t_{(jm,\a_j)}|\forall j,\a_j\}$.
All other conventions are as in previous papers${}^{12}$.

The action $\cs_1$
\bea
{\cal S}_1=\k S^-[g]+ \frac{1}{\p x} \int str\, A_{-}\left( J_+
-\frac \k 2 e_- -\frac \k 2 [\t,e_-]\right)
+ \frac{\k}{4\p x}\int str [\t,e_-]\del_-\t,
\label{actiongen}
\eea
with the affine currents $J_+=\frac \k 2 \del_+ g g^{-1}$, the gauge fields
$A_{-}\in\P_{+}\bar{g}$ and the ``auxiliary''
fields
$\t\in\P_{+1/2}\bar{g}$,
is invariant under the gauge transformations
\bea
g\rightarrow h g\qquad\quad
A_{-}\rightarrow\del_- h h^{-1}+h A_{-} h^{-1}\qquad\quad
\t\rightarrow \t + \P_{+\frac 1 2}\h,
\eea
where $h=\exp \h$, $\h\in\P_+\bar{g}$.

The gauge fields $A_{-}$ (Lagrange multipliers) impose the constraint
$\P_-J_+=\frac\k 2 e_-+\frac \k 2 [\t,e_-]$. Calling the constrained current
$J^c_+$, one performs the gauge transformation which brings $J_+^c$ in the form
$T+\frac \k 2 e_-$ where $T\in\ker ad \, e_+$, and obtain in this way the
fields
$T$ which are gauge invariant modulo the constraints, {\it i.e.} modulo the
equations of motion of the gauge fields $A_{-}$. They are of the form
$T\propto \P_{\ker ad\, e_+}J_++\cdots$
These currents are coupled to sources and the action is modified to
\bea
{\cal S}_2={\cal S}_1+\frac{1}{4\p x y} \int str \m T,\label{invact}
\eea
with the sources $\m\in \ker ad\, e_-$.
As the fields $T$ are only gauge invariant modulo terms proportional to
the equations of motion of the gauge fields, the resulting
non-invariance terms in $\d \cs_2$ are cancelled by modifying the
transformation rules for the gauge fields.
These modifications are proportional to the $\m$-fields and do not depend on
the gauge fields themselves. Because the gauge fields occur linearly in Eq.
(\ref{invact}), gauge invariance is restored.

In the next I will argue that the fields $T$ generate an extended Virasoro
algebra. A strong hint for this is the observation${}^{15,16}$ that
constraining a chiral WZW current as $J_+=T+\frac \k 2$ where $T\in\ker ad\,
e_+$, reduces the WZW Ward identities to the classical Ward identities of some
extended Virasoro algebra with currents $T=T^{(j,\a_j)}t_{(jj;\a_j)}$ and
$T^{(j,\a_j)}$ has conformal dimension $j+1$.

The functional $\G [\m]$
\bea
\exp -\G [\m]=\int [\d g g^{-1}][d\t][d A_{-}]\left(
\mbox{Vol}\left( \P_+\bar{g} \right) \right)^{-1}\exp-\left( {\cal S}_2
-\frac{1}{4\p x y} \int str \m\check{T}\right),\label{okok2}
\eea
is, if $T$ forms an extended Virasoro algebra, the induced action in the
light-cone gauge of the corresponding extended gravity theory.
The price  paid for modifying the transformation rule of the gaugefields $A_+$
is that the gauge algebra only closes on-shell. Such a system calls for the
Batalin-Vilkovisky formalism${}^{17}$ to gauge fix it. Introducing ghostfields
$c\in\P_+\bar{g}$ and anti-fields
$J^*_+\in\bar{g}$, $A^*_{-}\in\P_-\bar{g}$, $\t^*\in\P_{-1/2}\bar{g}$ and
$c^*\in\P_{-}\bar{g}$, the solution to the BV master equation is given by:
\bea
\cs_{\rm BV}&=&\cs_2-\frac{1}{2\p x}\int str c^*cc + \frac{1}{2\p x}\int str
J^*_+\left(\frac \k 2 \del_+ c + [c,J_+] \right)
+ \frac{1}{2\p x}\int str \t^*c\nonu
&&+\frac{1}{2\p x}\int str A^*_{-}\left(\del_- c + [c,A_{-}]
+\mbox{$\m$-dependent terms}\right).\label{bvbv}
\eea
The $\m$-dependent terms proportional to $A^*_{-}$ absorb all
complications arising from the non-invariance of $T$.

The gauge choice  $A_{-}=0$ is made by performing a canonical transformation
which changes $A^*_{-}$ into a
field,
the antighost $b\in\P_-\bar{g}$, and $A_{-}$ into an antifield $b^*$. The
gauge-fixed action reads:
\bea
{\cal S}_{\rm gf}=\k S^-[g]
+ \frac{\k}{4\p x}\int str [\t,e_-]\del_-\t
+\frac{1}{2\p x}\int str\, b\del_- c+\frac{1}{4\p x y} \int str \m \hat{T},
\label{actiongen2}
\eea
and the nilpotent BRST charge is:
\bea
Q=\frac{1}{4\p i x}\oint str\left\{ c \left( J_+ -\frac \k 2
e_--\frac \k 2[\t,e_-]+ \frac 1 2 J_+^{\rm gh}\right) \right\},
\eea
where $J_+^{\rm gh}=\frac 1 2 \{b,c\}$.

The only unknown in the action is the current $\hat{T}$. This reflects
the fact that I did not specify the explicit form of the $\m$ dependent terms
in Eq. (\ref{bvbv}). In order to
guarantee BRST invariance of the action
the currents $\hat{T}$ themselves have to be BRST invariant.
This determines them up to BRST exact pieces.

Following initial studies of this system${}^{18-20}$, the BRST cohomology of
$Q$ was solved in its full generality${}^{14}$ using spectral sequence
techniques${}^{21}$. I will omit most details here and just summarize the
results. The fields, $\F$, in the theory are assigned a double grading
$[\F]=(k,l)$,
$k,\,l\in\frac 1 2 {\bf Z}$, with $k+l\in {\bf Z}$ the ghostnumber:
$[J_+]=(m,-m)$ for $J_+\in\P_m\bar{g}$, $m\in\frac 1 2 {\bf Z}$, $[b]=(-m,m-1)$
for $b\in\P_{-m}\bar{g}$, $m>0$, $[c]=(m,-m+1)$ for $c\in\P_m\bar{g}$, $m>0$
and $[\t]=(0,0)$. The operator product expansions (OPE) are compatible with the
grading.
The BRST operator $Q$ is decomposed as $Q=Q_0+Q_1+Q_2$ where
\bea
Q_0=-\frac{\k}{8\p i x}\oint str c e_-\qquad
Q_1=-\frac{\k}{8\p i x}\oint str c\left[\t,e_-\right],\label{qsqr}
\eea
such that $[Q_0]=(1,0)$, $[Q_1]=(\frac 1 2,\frac 1 2)$ and $[Q_2]=(0,1)$. One
computes that
$Q_0^2=Q_2^2=\{Q_0,Q_1\}=\{Q_1,Q_2\}=Q_1^2+\{Q_0,Q_2\}=0$, but
$Q_1^2=-\{Q_0,Q_2\}=\frac{\k}{32 \p i x}\oint str \{ c [
\P_{1/2}c,e_-]\}$. A first fact is that, because of the existence of the
subcomplex with  trivial cohomology and generated by $\{ b,
\P_-\hat{J}_z-\frac \k 2 [ \t,e_-]\}$,
the full cohomology is isomorphic to the cohomology computed on the reduced
complex generated by $\{\P_{\geq 0}\hat{J}_+,\t,c\}$. I denoted the total
currents by $\hat{J}_+=J_++J_+^{\rm gh}$. Note that the OPEs also close on this
subcomplex.

Computing a spectral sequence, one can show${}^{14}$ that the cohomology is
generated by
$\hat{T}\equiv\sum_{j,\a_j}\hat{T}^{(j,\a_j)}t_{(jj;\a_j)} \in \ker ad \, e_+$
and $\hat{T}^{(j,\a_j)}$ has the form
\bea
\hat{T}^{(j,\a_j)}=\sum_{r=0}^{2j}\hat{T}^{(j,\a_j)}_r,
\eea
where $\hat{T}^{(j,\a_j)}_r$ has grading $(j-\frac r 2 , -j+\frac r 2)$. The
leading term
is of the form
\bea
\hat{T}^{(j,\a_j)}_0=C^j\left\{ \hat{J}_+^{(jj;\a_j)}+\frac \k 4
\sum_{\a_0}\d_{j,0}\d_{\a_j,\a_0} [\t ,[e_-,\t]]^{(00;\a_0)}
\right\},\label{lterm}
\eea
where the normalization constant $C$ will be fixed later on and the remaining
terms are recursively determined by a generalized tic-tac-toe construction:
\bea
Q_0\hat{T}^{(j,\a_j)}_r=-Q_1 \hat{T}^{(j,\a_j)}_{r-1} - Q_2
\hat{T}^{(j,\a_j)}_{r-2}.
\eea

The OPEs of $T^{(j,\a_j)}$ close modulo BRST exact terms. However, because
$T^{(j,\a_j)}$ has ghostnumber 0, a BRST exact term must be derived from a
ghostnumber $-1$ field. As the cohomology was computed on a reduced complex
which has no fields of negative ghostnumber, one concludes that the operator
algebra of $T^{(j,\a_j)}$ closes!

One can also show that the map $T^{(j,\a_j)}\rightarrow T^{(j,\a_j)}_{2j}$ is
an algebra isomorphism${}^{14,20}$. This is the so-called quantum Miura
transformation.

The only thing which remains to be shown is that the algebra is an extension of
the Virasoro algebra, {\it i.e.} it does contain the Virasoro algebra. One can
show that the energy-momentum tensor\footnote{\ninerm\baselineskip=11pt The
first term in Eq. (\ref{em11}) is $\hat{T}^{(1,0)}_0$, the second term
$\hat{T}^{(1,0)}_1$ and
the remainder forms $\hat{T}^{(1,0)}_2$. Requiring that this forms the Virasoro
algebra in the standard normalization fixes $C$:
$C=\frac{4y\k}{\sqrt{2}(\k+\tilde{h})}$.}:
\bea
\hat{T}^{\rm EM}&=&\frac{\k}{x(\k+\tilde{h})}\bigg( str \left\{\hat{J}_z e_-
\right\} + str \left\{ [\t,e_- ] \hat{J}_z\right\} +\frac 1 \k str
\left\{\P_0(\hat{J}_z)\P_0(\hat{J}_z) \right\} +\frac
{\k+\tilde{h}}{\k}str\left\{ e_-\del\hat{J}'_z\right\} \nonu
&& + \frac {1}{\k}str\left\{ \left[\P_0 (t^A),\left[\P_0 (t_A) ,\del\hat{J}'_z
\right]\right]e_-\right\}-\frac{\k+\tilde{h}}{4}str \left\{ [\t,e_-]\del\t
\right\}\bigg),\label{em11}
\eea
satisfies the Virasoro algebra with
\bea
c=\frac 1 2 c_{\rm crit} - \frac{(d_B-d_F)\tilde{h}}{\k+\tilde{h}} - 6 y (\k
+\tilde{h}),\label{cpretty}
\eea
where $c_{\rm crit}$ is the critical central extension of the algebra under
consideration,
$c_{\rm crit}=\sum_{j,\a_j}(-)^{(\a_j)}(12 j^2+12j+2)$,
$y$ is the index of the $\slt$ embedding, {\it i.e.} the ratio of the length
squared of the longest root of $\bar{g}$ with the length squared of the $\slt$
root, $d_B$, $d_F$ respectively, is the number of bosonic, fermionic
respectively, generators of $\bar{g}$ and $\tilde{h}$ is the dual Coxeter
number of $\bar{g}$. Adding a BRST exact term to Eq. (\ref{em11}), one gets the
energy-momentum tensor in the familiar KPZ form:
\bea
\hat{T}^{\rm IMP}
&\equiv&\frac{1}{x(\k+\tilde{h})} str J_+J_+ -\frac{1}{8xy}str e_0 \del_+ J_+
-\frac{\k}{4 x}tr\left( [\t,e_-]\del_+\t\right)\nonu
&&+\frac{1}{4x}strb[e_0,\del_+ c]-\frac {1}{2x}str b\del_+ c+\frac{1}{4x}str
\del_+ b
[e_0,c].\label{Timp}
\eea

A simple example is provided by the embeddings of $\slt$ in $sl(3)$. There are
two inequivalent embeddings of $\slt$ in $sl(3)$. For the first, the adjoint of
$\slt$ branches according to $\underline{8}\rightarrow [3]\oplus [5]$ which
gives rise to the $W_3$ algebra containing a dimension 2 current, the
energy-momentum tensor, and a conformal dimension 3 current. The other
embedding is characterized by $\underline{8}\rightarrow [1]\oplus [2]\oplus [2]
\oplus [3]$ and corresponds to the $W_3^{(2)}$ algebra containing the
energy-momentum tensor, two {\elevenit bosonic} dimension 3/2 currents and a
$U(1)$ current.

The effective action in the light-cone gauge, $W[\check{T}]$, of the
corresponding gravity theory${}^{12}$ is defined by
\bea
\exp -W[\check{T}]=\int [d\m ]exp-\left( \G [\m]
-\frac{1}{4\p x y} \int str \m\check{T}\right).\label{okok1}
\eea
where $\G [\m]$ was given in Eq. (\ref{okok2}). In order to compute the
effective action, one first feeds the information gained in the previous
analysis back into the solution of the BV master equation and one then chooses
a different gauge: $\t=\P_+[e_+,J_+]=0$. To achieve this, one makes a canonical
transformation in Eq. (\ref{bvbv}) which
interchanges fields and anti-fields for $\{\t,\t^*\}$ and $\{ \P_+[e_+,J_+],
\P_-[e_-,J^*_+]\}$. One finds
\bea
W[\check{T}]=\k_c S_-[g],\label{endresult}
\eea
where $\k_c=\k+2\tilde{h}$ and I used $[\d g g^{-1}]=[dJ_+]\exp \left(
-2\tilde{h} S^-[g]\right)$.
{}From Eq. (\ref{cpretty}) one gets the level as a function of the central
charge:
\bea
12 y \k_c=12 y\tilde{h}-\left(c-\frac 1 2 c_{\rm crit}\right)-
\sqrt{\left(c-\frac 1 2 c_{\rm crit}\right)^2- 24 (d_B-d_F)
\tilde{h}y}.\label{vv2}
\eea
Eq. (\ref{vv2}) provides an all-order expression for the
coupling constant renormalization.
The WZW model in Eq. (\ref{endresult}) is constrained by
\bea
\del_+ g g^{-1}+\frac{1}{4xy}str \left\{\P_{\rm NA}\left(\del_+ g
g^{-1}\right))\P_{\rm NA}\left(\del_+ g
g^{-1}\right)\right\}e_+=e_-+\frac{1}{\k
+\tilde{h}} \sum_{j,\a_j}
\frac{1}{2^{\frac 3 2 j-1}y^{j}}
\check{T}^{(j\a_j)} t_{(jj,\a_j)}
\eea
where  $\P_{\rm NA}\bar{g}$ is the projection on the centralizer of
$\slt$ in $\bar{g}$. I also used${}^{12,22}$ $J_+=\frac{\a_\k}{2}\del_+ g
g^{-1}$ with
$\a_\k=\k+\tilde{h}$.
\setcounter{equation}{0}
\setcounter{section}{5}
\vglue 0.5cm
{\elevenbf\noindent 5. Conclusions}
\vglue 0.4cm
\baselineskip=14pt
\elevenrm
I showed that one can associate an extended Virasoro algebra with  every
non-trivial embedding of $\slt$ in a semi-simple (super) Lie algebra. The
algebra is realized as a WZW model where a chiral, solvable group is
gauged---the gauge group being determined by the $\slt$ embedding. Algebras
that can be obtained in this way can be called ``simple'' extensions of the
Virasoro algebra. All other extended Virasoro algebras, can be called
``non-simple'' extended Virasoro algebras and I conjecture that they can be
obtained from ``simple'' ones by canonical manipulations such as adding free
fermions or $U(1)$ currents${}^{23}$, orbifolding${}^{24}$, further gauging of
affine subalgebras, etc. Relating the representation theory of the extended
Virasoro algebra with the representation theory of the underlying WZW model
remains a very interesting, open problem.

{}From Eq. (\ref{vv2}), one finds that
no renormalization of the coupling constant beyond one loop occurs if and only
if either $d_B=d_F$ or $\tilde{h}=0$ (or both). One gets $d_B=d_F$ for $su(m\pm
1|m)$, $osp(m|m)$ and
$osp(m+1|m)$ and $\tilde{h}=0$, for $su(m|m)$, $osp(m+2|m)$ and
$D(2,1,\alpha)$. This hints towards the existence of a generalized
non-renormalization theorem whose precise nature remains to be elucidated.

The close relation between this method of constructing extended Virasoro
algebras and extended 2D gravity theories, provides a good starting point for
the study of non-critical strings. Given an embedding of
$\slt$ in $\bar{g}$, one considers the corresponding $(p,q)$ minimal model as
the matter sector of the string theory. Its central charge $c_M$ is given by
Eq. (\ref{cpretty}), where $\k_{M}+\tilde{h}=p/q$. In order to cancel the
conformal anomaly, one needs to supplement the matter sector by a gauge sector
whose central charge $c_L$ is again given by eq, (\ref{cpretty}) but now
$\k_{L}+\tilde{h}=-p/q$. The corresponding $W$ string is now determined by
currents $T_{\rm tot}=T_{M} +  T_{L}$ and a BRST
charge $Q=\frac{1}{2\p i}\oint str c(T_{\rm tot} +\frac 1 2 T_{ghost})$,
where the ghost system contributes $-c_{\rm crit}$ to the central charge. In
order to explicitly perform this program, a covariant formulation is needed so
that one is not restricted to the light-cone gauge but that the conformal
gauge, which is more convenient,  can be used as well. This can again be
achieved using
WZW like techniques${}^{25}$.

A most challenging problem is the understanding of the geometry behind the
extensions of $d=2$ gravity. We saw that the Virasoro algebra appeared as the
algebra of residual symmetry after
gauge fixing a theory invariant under general coordinate transformations in
$d=2$. A similar statement for extensions of the Virasoro algebra remains to be
found.  Finally, a most exciting
application of the methods developed in this paper would be the study of
reductions of continuum algebras${}^{26}$ which will lead to
integrable theories in $d>2$! Work in these directions is in progress.

\setcounter{section}{5}
\vglue 0.5cm
{\elevenbf \noindent Acknowledgements \hfil}
\vglue 0.4cm
I would like to thank Hirosi Ooguri, Kareljan Schoutens, Kris Thielemans,
Walter Troost and Peter van Nieuwenhuizen, in collaboration with whom I
obtained some of the results presented here. I thank Jan de Boer, Gil Rivlis
and Tjark Tjin for interesting discussions and the organizers for an
interesting and very pleasant conference.

\vglue 0.5cm
{\elevenbf\noindent References \hfil}

\end{document}